\documentclass[12pt]{iopart}
\begin{document}

\article[Angle-resolved photoemission in high $T_c$ cuprates from theoretical viewpoints]
{topical}{Angle-resolved photoemission in high $T_c$ cuprates from theoretical viewpoints}

\author{Takami Tohyama\footnote[1]{E-mail: tohyama@imr.tohoku.ac.jp} and Sadamichi Maekawa\footnote[2]{E-mail: maekawa@imr.tohoku.ac.jp}}

\address{Institute for Materials Research, Tohoku University, Sendai 980-8577, Japan}

\begin{abstract}
The angle-resolved photoemission (ARPES) technique has been developed rapidly over the last decay, accompanied by the improvement of energy and momentum resolutions.  This technique has been established as the most powerful tool to investigate the high $T_c$ cuprate superconductors.  We review recent ARPES data on the cuprates from a theoretical point of view, with emphasis on the systematic evolution of the spectral weight near the momentum ($\pi$,0) from insulator to overdoped systems.  The effects of charge stripes on the ARPES spectra are also reviewed.  Some recent experimental and theoretical efforts to understand the superconducting state and the pseudogap phenomenon are discussed.
\end{abstract}

\pacs{79.60.-i, 74.20.-z, 74.25.Jb}

\submitted \SUST

\maketitle

\section{Introduction\label{Intro}}
Angle-resolved photoemission spectroscopy (ARPES) is a powerful technique to investigate the occupied electronic states below the Fermi level in solids.  The incident photons illuminate the surface of a single crystalline sample and then a photoelectron is ejected from the surface.  The kinetic energy and the angle of the outgoing electron are measured, from which the energy and momentum distributions of the occupied state are constructed \cite{Hufner}.

In the study of high $T_c$ superconductivity, ARPES has contributed greatly to the progress in understanding of the normal and superconducting properties \cite{ShenReview,RanderiaReview,TimuskReview}.  The improvement of the energy and momentum resolutions enables visualization of detailed features of low energy excitation near the Fermi level: they are the topology of the Fermi surface in Bi$_2$Sr$_2$CaCu$_2$O$_{8+x}$ (Bi2212) \cite{Olson1,Dessau1,Aebi,Ding1,Ding2,Saini1,Chuang,Feng}, in YBa$_2$Cu$_3$O$_{6+x}$ (YBCO) \cite{Campuzano1,Liu,Schabel1}, in La$_{2-x}$Sr$_x$CuO$_4$ (LSCO) \cite{Ino1}, and in electron-doped Nd$_{2-x}$Ce$_x$CuO$_4$ \cite{King1,AndersonR}, the superconducting gap \cite{Olson2,Dessau2,Hwu} and its anisotropic behavior \cite{Shen1,Ding3,Schabel2,Mesot,Gatt}, the pseudogap phenomenon in the normal state \cite{Loeser1,Ding4,Harris1,WhitePJ1,Loeser2,Marshall,Norman1,Norman2,Harris2}, the nature of quasiparticles in superconducting states of Bi2212 \cite{Norman3,Fedorov,Valla,Kaminski} and the effect of impurities and disorder \cite{Vobornik,WhitePJ2}.

The most dramatic features of high $T_c$ superconducting cuprates are that the undoped insulating antiferromagnets change into superconductors upon doping of carriers and that the superconductivity exists in a certain range of the carrier density.  ARPES experiments are also one of the primary sources of information of such changes of the electronic structure due to doping.  Systematic changes of the ARPES spectra with doping from the undoped insulators to underdoped and overdoped superconductors have been reported \cite{King2,LaRosa,Kim1,Campuzano2}, focusing mainly on the dispersion and spectrum at around momentum {\bf k}=($\pi$/$a$,0) or (0,$\pi$/$a$) where the superconducting gap with $d_{x^2-y^2}$ symmetry shows a maximum.  Here, $a$ is the Cu-Cu distance, which will be taken to be a unit length hereafter.  An interesting observation is that a broad spectrum at ($\pi$,0) in undoped samples \cite{Wells} is continuously evolved into a sharp peak in overdoped samples:  the spectrum in underdoped samples remains broad, and then it becomes sharper and sharper with further doping.  In addition, maximum positions of peaks along ($\pi$/2,$\pi$/2) to ($\pi$,0) show similar dispersion between undoped and underdoped samples, which follows a $d_{x^2-y^2}$ gap function \cite{Ronning}.  These results indicate a close relation between antiferromagnetism and superconductivity.  This issue will be reviewed in \sref{Insulator} and \sref{Doping}.

While the ($\pi$,0) spectra show interesting changes as a function of doping, the spectra along the (0,0)-($\pi$,$\pi$) direction reveal a sharp peak structure independent of doping in Bi2212 samples.  However, in contrast to Bi2212, LSCO samples show an extremely broad and weak intensity along the direction in the underdoped region \cite{Ino1,Ino2}.  The difference between the two typical families may be attributed to the presence of inhomogeneous charge distribution along the Cu-O direction, i.e., vertical charge stripes \cite{Tohyama2}.  \Sref{LSCO} will be devoted to this issue.

This review paper is mainly focused on the ARPES spectra and the underlying electronic states of high $T_c$ cuprates from the theoretical side.  One of the key points we would like to emphasize is the systematic change of the spectral weight with doping, in particular, near the momenta ($\pi$,0) and (0,$\pi$).  These momenta play a crucial role in the mechanism of superconductivity.  Therefore, investigating the spectral feature near ($\pi$,0) enables us to construct a model that can properly describe the physics of high $T_c$ cuprates and to clarify the mechanism of superconductivity.

This review is organized as follows.  In \sref{Model}, fundamental aspects of ARPES and the relation to the single-particle spectral function are briefly introduced.  A theoretical model Hamiltonian to study the ARPES spectra in this review is also introduced.  It is the $t$-$J$ model with second and third nearest neighbor hoppings, i.e., the $t$-$t'$-$t''$-$J$ model, which is widely used as an effective model to describe low energy excitations in the high $T_c$ cuprates.  \Sref{Insulator} is devoted to the ARPES spectrum for parent compounds, which represents the motion of a doped hole into two-dimensional antiferromagnetic insulators.  The physical implication of the ARPES data is discussed based on a spin-liquid picture obtained from theoretical analyses of the spectra.  The doping dependence of ARPES data for hole and electron dopings is shown in \sref{Doping}.  The evolution of the spectral weight at around ($\pi$,0) is emphasized comparing with theoretical results.  \Sref{LSCO} is devoted to the ARPES spectra of underdoped LSCO, which behave differently from those of Bi2212.  The spectral features are attributed to the presence of vertical stripes in this compound.  The issues of the superconducting and pseudogaps are discussed in \sref{Gap}.  In the superconducting state, ARPES data on Bi2212 show a quasiparticle peak followed by a dip and hump structure with higher binding energy as reviewed in \sref{GapExp}.  Since the peak/dip/hump feature is a direct consequence of the appearance of superconductivity, the explanation is crucial for the mechanism of high $T_c$ superconductivity.  The present state of experiments and several theoretical contributions are reviewed there.  A summary of this review is given in \sref{Summary}.

\section{ARPES and theoretical model\label{Model}}
\subsection{Spectral Function}

The two-dimensional nature of the CuO$_2$ plane in high $T_c$ cuprates is favorable to ARPES in determining the in-plane electronic states below the Fermi level, since the fact that the component of momentum parallel to the surface of samples is conserved between an outgoing photoelectron and a photohole created inside the samples makes it easy to map out the electronic states from the momentum and the kinetic energy of the photoelectron.  We note that, to obtain reliable data, it is necessary to be careful of the quality of the surface, since ARPES is a surface sensitive technique due to the short escape depth ($\sim$10\AA) of the outgoing electron.

For the full understanding of the measured energy distribution curve of the photoelectrons, or the ARPES intensity, one has to take into account a complicated process involving initial-state, final-state interactions and interactions between the outgoing electron and the surface and so on.  An effort to include all of the effects have been made based on the so-called one step model in which the absorption of the photon and the emission of electron are treated as a single event \cite{Bansil1}.  However, since the method relies on the band theory within local-density approximation, one can not properly involve the strong correlation effect that is of crucial importance for the CuO$_2$ plane.

Instead of the one-step calculation, we adopt the sudden approximation \cite{Hufner,Hedin} where final-state interactions are neglected.  The ARPES intensity is then given by
\begin{eqnarray}
I({\bf k},\omega)=I_0({\bf k})f(\omega)A({\bf k},\omega)\;,
\label{Ikw}
\end{eqnarray}
where {\bf k} is the momentum parallel to the CuO plane and $\omega$ is the excitation energy measured from the Fermi level.  $I_0({\bf k})$ is the squared matrix element that mainly comes from the condition of the dipole transition.   The dependence of the spectral weight on the polarization of the incident photon comes from this term.  $f(\omega)$ is the Fermi function: $f(\omega)=[\exp(\beta\omega)+1]^{-1}$, $\beta$ being the inverse temperature $1/k_{\rm B}T$.  $A({\bf k},\omega)$ is the spectral function that consists of two terms: $A({\bf k},\omega)=A_-({\bf k},\omega)+A_+({\bf k},\omega)$.  $A_-$ and $A_+$ are the electron-removal and additional spectral functions, respectively.  They are given by
\begin{eqnarray}
A_\pm({\bf k},\omega)&=&Z^{-1}\sum\limits_{n,m,\sigma}e^{-\beta E_m^N}
\left|\left< \Psi_n^{N\pm 1} \left| a_{{\bf k}\sigma} \right|\Psi_m^N \right>\right|^2
\delta(\omega\pm E_n^{N\pm 1} \mp E_m^N)\;,
\end{eqnarray}
where $a_{{\bf k}\sigma}=c_{{\bf k}\sigma}^\dagger$ and $c_{{\bf k}\sigma}$ for $A_+$ and $A_-$ respectively, $c_{{\bf k}\sigma}^\dagger$ ($c_{{\bf k}\sigma}$) being the creation (annihilation) operator of an electron with momentum {\bf k} and spin $\sigma$.  $\Psi_m^N$ is the wave function of the $m$th eigenstate with energy $E_m^N$ in the $N$-electron system.  $Z=\sum_m e^{-\beta E_m^N}$ is the partition function.  We note that $f(\omega)A({\bf k},\omega)=A_-({\bf k},\omega)$, being consistent with the fact that ARPES investigates the occupied states below the Fermi level.  The validity of the form of \eref{Ikw} in the high $T_c$ cuprates was discussed by Randeria {\it et al.} \cite{Randeria}.

\subsection{Model Hamiltonian}

To calculate the spectral function, we have to choose the Hamiltonian that describes the CuO$_2$ plane.  Due to strong Coulomb interaction on Cu ions and strong hybridization between Cu$d_{x^2-y^2}$ and O2$p_\sigma$ orbitals, a localized singlet state called the Zhang-Rice (ZR) singlet \cite{ZhangFC} is formed between a doped hole and a localized hole on a Cu ion.  The motion of the singlet and the interaction between the localized spins may be described by a $t$-$J$ model, given by
\begin{eqnarray}
H_{tJ}= -t\sum\limits_{\left<i,j\right>_{1{\rm st}},\sigma}
       (\tilde{c}_{i\sigma }^\dagger \tilde{c}_{j\sigma }+{\rm H.c.}) +
        J\sum\limits_{\left<i,j\right>_{1{\rm st}}}
      {{\bf S}_i}\cdot {\bf S}_j\;,
\label{HtJ}
\end{eqnarray}
where $\tilde{c}_{i\sigma}=c_{i\sigma}(1-n_{i-\sigma})$ is the annihilation operator of an electron with spin $\sigma$ at site $i$ with the constraint of no double occupancy, ${\bf S}_i$ is the spin operator and the summation $\left< i,j \right>_{1{\rm st}}$ runs over first nearest neighbor pairs.  Since the ZR singlet and localized spin exchange their positions by hopping process, the $t$ term corresponds to the hopping of the ZR singlet.  The $J$ term represents the exchange interaction between the localized spins.  The hopping process may exist between not only the first nearest neighbor pairs but also the second and third neighbor ones connecting sites at distance $\sqrt{2}a$ and $2a$, respectively.  The latter is given by
\begin{eqnarray}
H_{t't''}&=&-t'\sum\limits_{\left<i,j\right>_{2{\rm nd}} \sigma }
    c_{i\sigma }^\dagger c_{j\sigma }
     -t''\sum\limits_{\left<i,j\right>_{3{\rm rd}} \sigma }
    c_{i\sigma }^\dagger c_{j\sigma }+{\rm H.c.}\;,
\label{Ht't''}
\end{eqnarray}
$\left< i,j \right>_{2{\rm nd}}$ and $\left< i,j \right>_{3{\rm rd}}$
run over second and third nearest neighbor pairs, respectively.

The total Hamiltonian, $H_{tJ}+H_{t't''}$, is named the $t$-$t'$-$t''$-$J$ model.  The values of $J$ are known to be not strongly dependent on materials with the magnitude of 0.1-0.16~eV \cite{Bourges+Hayden,Mizuno}.  The ratio $J$/$t$ is evaluated to be around 0.4 from small cluster calculations in which Cu3$d_{x^2-y^2}$ and O2$p_\sigma$ orbitals are explicitly included \cite{Eskes,Hybertsen,Tohyama3}.  Therefore, $t$=0.35~eV when $J$=0.14~eV.  The importance of $t'$ along the plaquette diagonal was recognized from the analysis of the electronic structures of the two-dimensional cuprates \cite{Eskes,Hybertsen,Tohyama3}.  For hole-doped cuprates, $t'$ has been systematically found to be of negative sign in contrast to the nearest neighbor hopping amplitude $t$ with positive sign, and of about 20\% to 40\% of its magnitude.  On the other hand, electron-doped cuprates need opposite signs for all hopping amplitude \cite{Tohyama4,Gooding}.  The presence of the third neighbor hopping $t''$ was shown by a perturbative treatment of the Cu-O Hamiltonian \cite{Matsukawa}.  However, the determination of precise values of the hopping parameters from the analysis of the electronic structures is very difficult due to the complicated process contributing to them.  Instead, the use of the Fermi surface observed experimentally or obtained by the band structure calculations may be effective in evaluating $t'$ and $t''$ \cite{Tanamoto}.  \Fref{FSexp} represents the Fermi surface of Bi$_2$Sr$_2$CaCu$_2$O$_{8+x}$ (Bi2212) \cite{Ding2} and of La$_{2-x}$Sr$_x$CuO$_4$ (LSCO) \cite{Ino1} determined by ARPES experiments.  The ratios $t'$/$t$ and $t''$/$t$ are obtained by fitting the tight binding ($t$-$t'$-$t''$) Fermi surface to the experimental ones.  They are $t'$/$t$=$-$0.34 and $t''$/$t$=0.23 for Bi2212 and $t'$/$t$=$-$0.12 and $t''$/$t$=0.08 for LSCO.  The resulting tight binding Fermi surfaces are shown in \fref{FStheory}.  The fitting is not fine but enough for the characteristics of the effect of $t'$ and $t''$ to be included.  The difference of the Fermi surface between Bi2212 and LSCO is attributed to the contribution of apical oxygen to the band dispersion \cite{Feiner}.

Finally, we would like to emphasize that much care is necessary when the spectral intensities in the $t$-$t'$-$t''$-$J$ model are quantitatively compared with ARPES data.  This is due to the fact that the photoemission cross section depends on the orbitals from which electrons originate: the cross section of O2$p$ is much larger than that of Cu3$d$ when photon energy is about $h\nu$=20~eV \cite{Kim4}.  Since the $t$-$t'$-$t''$-$J$ model contains information on both copper and oxygen wavefunctions, it is necessary to take additional phase factors into account in performing a quantitative comparison of the intensity \cite{Eroles}.

\section{Single hole in antiferromagnetic insulator\label{Insulator}}

In this section, the ARPES spectrum for parent compounds is analyzed in terms of the motion of a doped hole into two-dimensional antiferromagnetic insulators.  The role of the long-range hoppings of the hole is examined, and novel spin-liquid pictures are proposed as the physical implication of the ARPES data.

\subsection{Dispersion and line shape}

The first ARPES experiment on undoped compounds was performed by Wells {\it et al.} \cite{Wells} for Sr$_2$CuO$_2$Cl$_2$.  The data were taken at the temperature of 350~K above the N\'eel temperature of 256~K.  Kim {\it et al.} \cite{Kim1} found that the dispersion is similar to that at 150~K well below the N\'eel temperature, although the spectral intensity slightly depends on temperature.  \Fref{SCOCspectrum} shows the ARPES data by Kim {\it et al.} \cite{Kim1}.  Along the (0,0)-($\pi$,$\pi$) direction, the low energy peak, referred to as the quasiparticle peak, moves by about 0.3~eV toward the low binding energy side, reaches its minimum at ($\pi$/2,$\pi$/2), and then folds back with a rapid reduction of the intensity.  Along the (0,0)-($\pi$,0) cut, the spectral weight has its maximum at around (2$\pi$/3,0), and as {\bf k} is increased further toward ($\pi$,0), the low energy features are suppressed.  The dispersion of the peak is illustrated in \fref{SCOCdispersion}.

The ARPES on the undoped compound represents the motion of a doped hole in the two-dimensional antiferromagnetic insulator that has extensively been studied by using the $t$-$J$ model for more than ten years since the discovery of the high $T_c$ superconductors \cite{Dagotto}.  From the comparison between the ARPES data and the $t$-$J$ results \cite{Wells}, it is found that the $t$-$J$ model can quantitatively explain the observed dispersion from (0,0) to ($\pi$,$\pi$) with a width of 2.2$J$.  However, the model does not explain the dispersion with a width of about 2$J$ along the (0,$\pi$)-($\pi$,0) direction: it predicts a nearly flat dispersion along this direction given by an approximate form that has the reduction of energy scale from $t$ to $J$,
\begin{eqnarray}
E_0({\bf k})=-0.55J\left( \cos k_x + \cos k_y \right)^2\;,
\label{tJDispersion}
\end{eqnarray}
where the energy is measured from the ($\pi$/2,$\pi$/2) point, and the coefficient 0.55$J$ is determined so as to give the band width of 2.2$J$.  This discrepancy has led to intense theoretical studies, from which it was found that the discrepancy may be resolved by introducing hopping to second and third nearest neighbors, $H_{t't''}$, shown in \eref{Ht't''} \cite{Nazarenko,Kyung1,Xiang,Belinicher,Eder,LeeTK,Lema,Leung,Sushkov}.  In regions where AF correlations are important, $t'$ and $t''$ are not severely renormalized due to hopping in the same sublattice.  Therefore, the dispersion may be approximately written as
\begin{eqnarray}
E_1({\bf k})=-4t'_{eff}\cos k_x\cos k_y-2t''_{eff}\left( \cos 2k_x + \cos 2k_y \right)\;,
\label{t't''Dispersion}
\end{eqnarray}
where $t'_{eff}$ and $t''_{eff}$ are effective values roughly proportional to $t'$ and $t''$, respectively.  This strongly affects the energy at ($\pi$,0), lowering it by 4($t'_{eff}-t''_{eff}$) ($t'_{eff}<$0 and $t''_{eff}>$0).  More precise calculations of the dispersion in the $t$-$t'$-$t''$-$J$ model are performed by using a self-consistent Born approximation \cite{Schmitt,Kane,Martinez}.  The values of $t'$ and $t''$ are expected to be the same as those for Bi2212, because band calculations show similar Fermi surface topologies for Sr$_2$CuO$_2$Cl$_2$ \cite{AndersonOK} and Bi2212 \cite{Massidda}.  The calculated dispersion and a fitted curve to \eref{tJDispersion} and \eref{t't''Dispersion} are shown in \fref{SCOCdispersion}, and reproduce the experimental ones.

While the dispersion of the lowest energy peak (quasiparticle peak) helps to characterize the motion of a hole, the ARPES line shape also contains important information.  \Fref{ARPESweight}(a) shows the Sr$_2$CuO$_2$Cl$_2$ spectra at ($\pi$/2,$\pi$/2), ($\pi$/2,0), and ($\pi$,0) \cite{Kim1}.  The quasiparticle peak at ($\pi$/2,$\pi$/2) is sharp, while at the ($\pi$,0) point the peak is strongly suppressed.  Such a suppression is reproduced very well by the $t$-$t'$-$t''$-$J$ model but not by the $t$-$J$ model as shown in \fref{Akw1hole} \cite{Kim1}.  The exact diagonalization technique  \cite{Dagotto} for a 4$\times$4 cluster with periodic boundary condition is used for the calculation of the spectral function at $T$=0.  The fact that the $t'$ and $t''$ terms are responsible for the suppression can be understood as follows:  For the chosen values of $t'$ and $t''$, the energy of the N\'eel state with a hole, $|\Psi_{\bf k}^{N\acute{e}el}>$ is relatively low for ${\bf k}=(\pi,0)$, reducing the one-hole-N\'eel-state character of the final state associated with the quasiparticle peak.  Since the initial state is antiferromagnetically ordered, the photoemission spectral weight is expected to be proportional to the one-hole-N\'eel-state character of the final state, ($\left|\left<f\left| c_{{\bf k}\sigma} \right|i\right>\right|^2= \left|\left<f\left| c_{{\bf k}\sigma} \right|AF\right>\right|^2\sim \left|\left<f|\Psi_{\bf k}^{N\acute{e}el}\right>\right|^2$).  The vertical bars in \fref{Akw1hole}(b) denote the calculated weight of $\left|\Psi_{\bf k}^{N\acute{e}el}\right>$ in the eigenstate corresponding to the quasiparticle.  The fact that the feature at ($\pi,0$) in the $t$-$t'$-$t''$-$J$ calculation is weaker than the corresponding one in the $t$-$J$ calculation means that for a hole with ${\bf k}=(\pi,0)$ the interplay of $t'$ and $t''$ with $t$ and $J$ causes a further weakening of the antiferromagnetic spin correlations.  The implications of this will be discussed below.

Another interesting feature is the presence of a shoulder separated by about 0.4~eV below the primary peak along the (0,0)-($\pi$,0) direction as denoted by bars in \fref{SCOCspectrum}(b).  The second structure is also reproduced by the $t$-$t'$-$t''$-$J$ model.  This structure may be due to string resonance states in the structure coming from the incoherent motion of the hole \cite{Dagotto}.

\Fref{SCOCspectrum}(a) also shows that the ($\pi$/2,0) spectrum is broader than the ($\pi$/2,$\pi$/2) one.  The $t$-$t'$-$t''$-$J$ model, however, can not explain this: the ($\pi$/2,0) spectrum is as sharp as the ($\pi$/2,$\pi$/2) one.  This discrepancy is resolved by introducing an additional exchange interaction $J'$ along the plaquette diagonal that induces the magnetic frustration in the spin background \cite{Shibata}.

\subsection{Spin liquid states and spectral function}

As discussed above, the dispersion from ($\pi$/2,$\pi$/2) to ($\pi$,0) requires additional terms to the $t$-$J$ model.  The spectral line shape at ($\pi$,0) is also different from the model.  Along this direction, Ronning {\it et al.} \cite{Ronning} have shown that the dispersion in another insulating compound Ca$_2$CuO$_2$Cl$_2$ follows the form of a $d_{x^2-y^2}$ gap as shown in \fref{CCOCdwave}(a).  This form is consistent with that of underdoped compounds.  In addition, the line shape of the ($\pi$,0) spectrum is continuously evolved from insulating to underdoped samples as seen in \fref{CCOCdwave}(b).  These data clearly demonstrate the continuity of the electronic states from the antiferromagnetic insulators to the superconductors, and have a great impact on theoretical studies of the relation between antiferromagnetism and superconductivity \cite{AndersonPW,Affleck,Laughlin,Wen,ZhangSC}.

If the dispersion completely follows the $d_{x^2-y^2}$-gap function that is of linear function in the plot of \fref{CCOCdwave}(a), it means that the dispersion starts linearly from the ($\pi$/2,$\pi$/2) point.  Such a behavior is obtained when a spin-liquid state or $d$-wave resonant valence bond (RVB) state is taken as the ground state of the Heisenberg model and the separation of spin and charge degrees of freedom, i.e., the presence of spinons and holons, is assumed \cite{Laughlin,Wen}.  Then, the quasiparticle dispersion of the single hole corresponds to the dispersion of the spinon given by \cite{Laughlin}
\begin{eqnarray}
E_{spinon}({\bf k})=1.6J\sqrt{\cos^2k_x + \cos^2k_y}\;.
\label{SpinonDispersion}
\end{eqnarray}
This dispersion is also plotted in \fref{SCOCdispersion}.  A similar linear dispersion near the ($\pi$/2,$\pi$/2) point was reported by Weng {\it et al.} \cite{Weng} introducing the effect of the phase string into the single-hole spectral function, and also by Hanke {\it et al.} \cite{Hanke} using a modified SO(5) theory \cite{ZhangSC} that unifies antiferromagnetism and $d$-wave superconductivity explicitly taking the Mott-Hubbard gap into account.

However, the data on Ca$_2$CuO$_2$Cl$_2$ shown in \fref{CCOCdwave}(a) seem to deviate from a linear function near {\bf k}=($\pi$/2,$\pi$/2).  Rather, they seem to fit a function with a quadratic term near the ($\pi$/2,$\pi$/2) point \cite{Kim2}.  This is consistent with a dispersion obtained from the self-consistent Born approximation of the $t$-$t'$-$t''$-$J$ model shown in \fref{SCOCdispersion}, where the antiferromagnetic long-range order is assumed in the ground state.  Although the model describes the observed dispersion very well, a frequently arising question concerns the applicability of $t'$ and $t''$ terms on the real system: the experimental fact that the dispersion in the (0,$\pi$)-($\pi$,0) direction is almost the same as in the (0,0)-($\pi$,$\pi$) direction is hard to be ascribed to the $t'$ and $t''$ terms, because the (0,0)-($\pi$,$\pi$) dispersion is regulated by $J$ \cite{Laughlin,Hanke}.  This question is resolved by examining the dependence of the quasiparticle energy difference $E(\pi,0)-E(\pi/2,\pi/2)$ on the values of $t'$ and $t''$ as well as on $J$ as shown in \fref{BandWidth} \cite{Tohyama1}.  The difference in the $t'$-$t'$-$t''$-$J$ model has a tendency to be saturated around $\alpha$$\sim$1, $\alpha$ being a scaling factor of $t'$ and $t''$ as $t'$($t''$)=$-$0.34(0.23)$\alpha t$.  At $\alpha\sim 1$, the energy difference is about 0.8$t$=2$J$.  This number is of importance because the incoherent spectrum of a hole starts from the energy position of about 2$J$ above the quasiparticle peak at ($\pi$/2,$\pi$/2) (see the insets in \fref{AkwSqwNqw}).  This means a crucial role of the incoherent motion of the hole.  Since the position of the incoherent structure is regulated by $J$, $E(\pi,0)-E(\pi/2,\pi/2)$ has the same $J$ dependence as $E(0,0)-E(\pi/2,\pi/2)$, at least, in the realistic range $J$/$t$=0.2-0.6, as seen in \fref{BandWidth}(b).  Therefore, the dispersion of the hole isotropic around ($\pi$/2,$\pi$/2) is governed by the spin degree of freedom, similar to that from the $d$-wave RVB picture \cite{Affleck,Laughlin,Wen}.  From examinations of the spin correlation around a doped hole, it was concluded that a novel spin-liquid state is realized around the hole with momentum ${\bf k}$=($\pi$,0) in contrast to a N\'eel-like state at around ($\pi$/2,$\pi$/2) when the values of $t'$ and $t''$ are large enough to reproduce the observed dispersion \cite{Tohyama1,Martins}.  Note that this spin-liquid state should be distinguished from the $d$-wave RVB state \cite{Affleck,Laughlin,Wen}: the former is seen in the one-hole state with {\bf k}=($\pi$,0) and the excitation energy of $\sim$2$J$, while the RVB theory predicts a spin-liquid state which is independent of the momentum of a doped hole.

The dynamical properties of spin and charge degrees of freedom also provide us useful information on the quasiparticle state with ($\pi$,0).  \Fref{AkwSqwNqw} shows the dynamical spin and charge correlation functions, $S({\bf q},\omega)$ and $N({\bf q},\omega)$ \cite{Tohyama5}, on a $\sqrt{20}$$\times$$\sqrt{20}$ cluster with one hole and the single-hole spectral function at {\bf k}=($\pi$,0) obtained by using the same cluster \cite{Tohyama1}.  In the $t$-$J$ model, both spin and charge components are involved in the quasiparticle state, while only the spin component remains in the $t$-$t'$-$t''$-$J$ model.  This is because of the separation of the spin and charge degrees of freedom, being consistent with the spin-liquid picture mentioned above.  However, the separation is incomplete unlike the 1D case of the $t$-$J$ model in which spin and charge are separated \cite{Kim3}.  Note that the spin-charge separation in the one-dimensional cuprates is found in ARPES data of SrCuO$_2$ \cite{Kim3} and Sr$_2$CuO$_3$ \cite{Fujisawa}.

Since the motion of a doped hole in the antiferromagnetic Mott insulator is the first step toward the understanding of novel physical properties in high $T_c$ cuprates, more complete understanding of the single-hole spectral properties in the whole excitation-energy range is necessary in the future.  For example, the similarity and difference of the ARPES spectra among the one-, two-dimensional, and ladder \cite{Takahashi} compounds may provide us crucial information on this subject.

\section{Doping dependence of ARPES line shape\label{Doping}}

In this section, we discuss the systematic evolution of the line shape of the ARPES spectra upon carrier doping in the normal state above $T_c$.  Both hole and electron dopings are shown.

\subsection{Hole doping}

\Fref{ARPESweight}(b) and \fref{ARPESweight}(c) show the doping dependence of the ARPES spectra at $T$=100~K above $T_c$ \cite{Kim1}.  As is the case of undoped Sr$_2$CuO$_2$Cl$_2$, in the underdoped Bi2212  the peak at the (0,0)-($\pi$,$\pi$) Fermi crossing is sharp while the peak at ($\pi$,0) is broad and suppressed.  In the overdoped sample, the peak at ($\pi,0$) moves closer to the Fermi energy and becomes sharp with increased intensity while the line shape on the (0,0)-($\pi$,$\pi$) cut remains more or less the same as in the underdoped sample \cite{Marshall,Kim1}.  The shift of the broad quasiparticle peak at ($\pi$,0) toward the Fermi energy is also seen in \fref{CCOCdwave}.  A similar doping dependence was observed in single plane Bi$_2$Sr$_{2-x}$La$_x$CuO$_{6+\delta}$ (Bi2201) \cite{Harris2}, although the line shape on the (0,0)-($\pi$,$\pi$) cut in the underdoped Bi2201 is less sharp.  \Fref{DopingDispersion} shows the spectral peak centroids against {\bf k} along high symmetry lines for Bi2212 with various hole densities \cite{Marshall}.  This clearly shows that, as the hole doping is reduced, the peak position around the ($\pi$,0) point shifts to higher binding energy, while the features near ($\pi$/2,$\pi$/2) are not affected.  The gap near ($\pi$,0) in the underdoped samples is sometimes called {\it high energy pseudogap}, the energy of which is about 100 to 200~meV ($\sim$J).

The $t$-$t'$-$t''$-$J$ model reproduces the experimental evolution of the quasiparticle peaks on hole doping \cite{Kim1,Eder}.  \Fref{Akwdoping}(a) and \fref{Akwdoping}(b) show the $t$-$t'$-$t''$-$J$ results of the spectral function for a $\sqrt{20}\times\sqrt{20}$ cluster \cite{Kim1}.  The spectrum at ($\pi$/5,3$\pi$/5) can be compared with the experimental data at the point $\alpha$ for the underdoped case in \fref{ARPESweight}(b).  One sees that the theoretical peak at ($\pi$/5,3$\pi$/5) is sharp while the spectrum at ($\pi$,0) is broad.  In the overdoped case with doping concentration of $\delta$=0.3, the peak at ($\pi$/5,3$\pi$/5) moves above the Fermi energy and the peak at (2$\pi$/5,$\pi$/5) becomes sharp.  The ($\pi$,0) spectrum at $\delta$=0.3 also shows a sharp feature at the lowest excitation energy with increased spectral weight compared to $\delta$=0.1.  This behavior is consistent with the experimental data mentioned above.  We note that the peak position of the ($\pi$,0) spectra shifts to near the Fermi level with doping \cite{Eder}.  Also a flat region of the band around the ($\pi$,0) point is obtained by using the exact diagonalization together with analytical method based on a spin-polaron picture by Yin {\it et al.} \cite{Yin}, being consistent with that of \fref{DopingDispersion}.

The great breadth of the quasiparticle peak at ($\pi$,0) for $\delta$=0.1 is caused by the reduction of the quasiparticle weight, which may come from two alternative sources.  (i) The same mechanism of the coupling between charge motion and spin background discussed above for the undoped case could be still effective in the lightly doped material: a spin-liquid state around a hole with momentum ($\pi$,0) induced by $t'$ and $t''$ survives even in the underdoped region.  This may be a probable explanation of the origin of the high energy pseudogap \cite{Tohyama1}. (ii) Alternatively, there could be a larger phase space for decay of quasiparticles because of strong coupling of the photo-hole to collective magnetic excitations near {\bf q}=($\pi$,$\pi$) as proposed by Shen and Schrieffer \cite{Shen2}. The Fermi surface topology is changed by the inclusion of $t'$ and $t''$ in a way that enhances this coupling.  Both of these mechanisms are likely to have less importance for the overdoped case.

A qualitatively similar evolution of the spectral function upon doping is obtained for the Hubbard model with $t'$ by using the quantum Monte Carlo (QMC) technique supplemented by the maximum-entropy method (MEM) \cite{Duffy}.  \Fref{QMC094} shows the spectral function of the $t$-$t'$-$U$ Hubbard model on a 6$\times$6 cluster with $\delta$=0.06.  Although the temperature is nearly as high as the value of $J$, a broad line shape at ($\pi$,0) and a sharp peak at around ($\pi$/2,$\pi$/2) are seen, being consistent with the $t$-$t'$-$t''$-$J$ results.  An analytical result at low temperatures is given by Kyung \cite{Kyung2}.  For the Hubbard model without $t'$, the doping dependence of the spectral function was studied by Preuss {\it et al.} \cite{Preuss} based on QMC+MEM, where the effects of antiferromagnetic spin correlations in the underdoped regime on the spectral function are emphasized to explain the ARPES data.  The flat dispersion near ($\pi$,0) was argued by Tsunetsugu and Imada \cite{Tsunetsugu} in connection with $d$-wave superconducting correlations, where not only $t'$ and $t''$ but also pair-hopping process \cite{Assaad} plays an important role.

Without the introduction of $t'$ and $t''$, a gauge theory based on the spin-charge separation scenario explains the spectral line shape and its doping dependence at around ($\pi$,0) and ($\pi$/2,$\pi$/2).  Dai and Su \cite{DaiX} calculated the spectral function by introducing the staggered gauge field fluctuation as well as the long wave gauge field fluctuation into a U(1) mean filed theory.  \Fref{U(1)+SU(2)}(a) shows the results for the underdoped and overdoped cases.  In the underdoped case, where the staggered gauge fluctuation is important, the line shape is very broad at ($\pi$,0) while sharp at ($\pi$/2,$\pi$/2).  In the overdoped case, the spectral function behaves differently, since the staggered gauge fluctuation becomes less important.  The latter case resembles a situation studied by Lee and Nagaosa \cite{LeePA}.  To overcome the defects of the U(1) mean filed theory, Wen and Lee \cite{Wen} proposed an SU(2) theory in which the SU(2) symmetry is preserved away from half filling by introducing two kinds of slave bosons.  A flat band near the ($\pi$,0) point similar to the experiments is obtained in the underdoped case as shown in \fref{U(1)+SU(2)}(b).  In this picture, the energies at the ($\pi$,0) and (0,0) points are degenerate at half filling according to \eref{SpinonDispersion}.

A phenomenological analysis of ARPES line shape for Bi2212 was performed by Misra {\it et al.} \cite{Misra}.  They found that the intensity at ($\pi$,0) changes from a non-Fermi liquid form $I(\omega)\propto\omega^{-1/2}$ \cite{Chubukov1} in the underdoped regime to a Fermi liquid form $I(\omega)\propto(\omega^2+\Gamma^2)^{-1}$ in the overdoped regime, $\Gamma$ being a cutoff frequency, while the intensity near the ($\pi$/2,$\pi$/2) point is fitted by either the Fermi liquid or non-Fermi liquid form for underdoping and by the Fermi liquid one for overdoping.  This implies that the broad line shape at ($\pi$,0) contains a clue of anomalous normal state properties in the underdoped regime.

\subsection{Electron doping}

The ARPES data in electron-doped Nd$_{2-x}$Ce$_x$CuO$_4$ (NCCO) were reported \cite{AndersonR} for optimal and slightly overdoped samples \cite{King1,AndersonR}.  The observed Fermi surface topology is similar to that of Bi2212 \cite{King1}.  However, the spectral line shape and its momentum dependence are different as seen in \fref{ARPESweight}(d).  The peak at ($\pi$,0) is at considerably higher excitation energy than that for Bi2212, although the experimental resolution is not enough to reveal details of the line shape.

The spectral function of the $t$-$t'$-$t''$-$J$ model for electron doping is qualitatively different from those for hole doping.  \Fref{Akwdoping}(c) shows that the quasiparticle spectral weight at ($\pi$,0) is large compared to the hole-doped case.  For electron doping, $t'$ appears to enhance the antiferromagnetic correlations near half filling due to its positive sign \cite{Tohyama4,Gooding}.  Therefore, the N\'eel-like state becomes more stable, unlike the hole-doped case, resulting in large quasiparticle weight.  Also note that the binding energy of the ($\pi$,0) peak is larger than that for the hole-doped case.  This is due to a band structure effect that makes the energy of the band near ($\pi$,0) higher in contrast to the hole-doped case where the band near ($\pi$,0) is flat.  Very recently, a precise measurement of the dispersion and line shape was performed on the superconducting samples \cite{Armitage}.  Two distinct components are observed in the data.  One is a dispersion with higher binding energy, which resembles to the dispersion of the insulators.  The other is a dispersion near the Fermi level with small width.  The higher binding-energy feature seems to be consistent with the $t$-$t'$-$t''$-$J$ results shown in \fref{Akwdoping}(c), while the origin of the low energy feature remains to be solved.

Recent optical measurements show an anomalous temperature-dependent spectral feature accompanied by gap-like structures for oxygenated non-superconducting samples \cite{Onose}.  It is highly desirable to construct a theoretical model that explains the optical, ARPES data and also the symmetry of superconducting state for which the evidence of an $s$-wave is accumulating.  This will clarify the similarity and difference between $p$- and $n$-type superconductors.

\section{Effect of charge stripes on LSCO\label{LSCO}}

So far, we have shown the data for Bi2212 system as hole-doped samples.  La$_{2-x}$Sr$_x$CuO$_4$ (LSCO) is another typical system that is widely examined to investigate the mechanism of superconductivity \cite{Kastner}.  It has a simple crystal structure with a single CuO$_2$ layer and the hole density in the CuO$_2$ plane is changeable in a wide range from $x$=0 to 0.35.

Recently, Ino {\it et al.} \cite{Ino1} performed the ARPES experiment on LSCO and reported different behaviors from Bi2212. (i) {\it Fermi surface topology} [see \fref{FSexp}(b)]: The Fermi surface undergoes a dramatic change from a hole-like shape centered at {\bf k}=($\pi$,$\pi$) in underdoped and optimally doped samples to an electron-like one centered at (0,0) in overdoped ones.  (ii) {\it Spectral line shape} (see \fref{LSCOspectrum}): In the underdoped case of $x$=0.1, the spectrum near ($\pi$/2,$\pi$/2) along the (0,0)-($\pi$,$\pi$) direction is very broad and weak in contrast to the case of underdoped Bi2212 [see \fref{Akwdoping}(b)] where a sharp peak appears near ($\pi$/2,$\pi$/2).

Not only ARPES but also other experimental data in LSCO also revealed anomalous behaviors of structural \cite{Saini2,Moodenbaugh,Bozin}, electronic \cite{Ino3,Uchida} and magnetic properties \cite{Yamada}.  In particular, at $x$=0.12, the incommensurate antiferromagnetic long-range order exists \cite{Suzuki}.  Similar antiferromagnetic order was observed in the low temperature tetragonal (LTT) phase of Nd-doped LSCO, La$_{1.48}$Nd$_{0.4}$Sr$_{0.12}$CuO$_4$, accompanied by charge order.  This is interpreted as charge/spin stripe order that consists of vertical charge stripes and antiphase spin domains \cite{Tranquada}.

From the theoretical side, disorder or fluctuation of stripe phases has been argued as essential physics of high $T_c$ superconductors \cite{Emery1,Zaanen,Castellani}.  However, it is controversial whether the $t$-$J$ model itself has the stripe-type ground state \cite{WhiteSR1,Hellberg,Kobayashi}.  Furthermore, the long-range hopping matrix elements weaken the stripe stability \cite{Tohyama6,WhiteSR2}.  A possible origin of the appearance of stable stripe phase is due to the presence of the long-range part of the Coulomb interaction \cite{Emery2,Seibold1} and/or the coupling to lattice distortions.  In LSCO, the LTT fluctuation may help the latter mechanism \cite{Saini2,Bozin}.

Tohyama {\it et al.} \cite{Tohyama2} introduced a stripe potential into a $\sqrt{18}\times\sqrt{18}$ cluster of the $t$-$t'$-$t''$-$J$ model, which makes the vertical stripes stable.  The values of $t'$ and $t''$ are the same as those discussed in \sref{Model}, and two holes are involved in the ground state to simulate an underdoped system.  The stripe potential $V_s$ is assumed to depend on the number of holes $n_h$ in each column of the cluster, that is, $V_s(n_h)$ with $V_s(0)=V_s(1)=0$, $V_s(2)=-2V$, and $V_s(3)=-3V$, with $V$$>$0.  The $x$ and $y$ directions are assigned to be perpendicular and parallel to the stripes, respectively.

\Fref{ARPESstripe} shows the spectral function of the $t$-$t'$-$t''$-$J$ model with the stripe potential \cite{Tohyama2}.  When there is no potential [\fref{ARPESstripe}(a)], the quasiparticle peaks with large weight are clearly seen below and above the Fermi level at ($\pi$/3,$\pi$/3) and (2$\pi$/3,2$\pi$/3), respectively.  For $V$/$t$=1 [\fref{ARPESstripe}(b)], however, there is no distinct quasiparticle peak and the spectra become more incoherent.  This means that the stripe has a tendency to make the spectrum broad along the (0,0)-($\pi$,$\pi$) direction, which is consistent with the ARPES data \cite{Ino1} shown in \fref{LSCOspectrum}.

For $V$/$t$=2 [\fref{ARPESstripe}(c)], split-off states from incoherent structures are seen at $\omega$/$t$=0.7 for ($\pi$/3,$\pi$/3) and (2$\pi$/3,0).  This is due to localization of carriers along the direction perpendicular to the stripes.  For $V$/$t$=1, the tendency to the localization may be involved in the low energy peaks at (2$\pi$/3,0), but can not be recognized at ($\pi$/3,$\pi$/3).  Such a localized feature near ($\pi$,0) was also obtained as a flat structure of the energy dispersion by Salkola {\it et al.} \cite{Salkola}.  They use a phenomenological approach, where the vertical stripe is introduced into a tight binding model as a spin-dependent potential, which ensures the existence of the antiphase spin domain, and a random distribution of the stripes are taken into account.  \Fref{StripeBand} shows the resulting dispersion.  Similar dispersion is reported in models that contain vertical stripes as a mean-filed solution of the $t$-$t'$-$U$ Hubbard model \cite{Machida} and as a result of incommensurate charge-density scattering \cite{Seibold2}.  Note that such flat bands along the (0,0)-($\pi$,0) direction are observed LSCO and Nd-doped LSCO \cite{Zhou} together with other materials \cite{Balatsky}.

In the insulating regime whose carrier density is below the critical concentration for superconductivity of $x$=0.05, not vertical stripes but diagonal ones have been reported by neutron scattering experiments \cite{Wakimoto}.  Therefore, the direction of stripes is also of importance for understanding of the electronic states of LSCO.  In fact, enhanced ARPES spectral weights along the (0,0)-($\pi$,$\pi$) direction were observed at and below the critical concentration of $x$=0.05 in contrast to the broad ones in the underdoped regime \cite{Ino2}.  A diagonal stripe potential is introduced into a small $t$-$t'$-$t''$-$J$ cluster as for the case of vertical stripes \cite{Tohyama2}.  It is found that, even if the stripe potential is turned on, the ($\pi$/2,$\pi$/2) spectrum remains sharp, which is consistent with the experiment.  This is because the potential does not change the nature of diagonal hole-pair configuration dominant in the ground state.

These consistent results between experiments and theories suggest that the concept of stripes is an essential ingredient for the explanation of the physical properties of LSCO, although the $d$-wave superconductivity is suppressed by the stripes \cite{Tohyama2}.

\section{Pseudogap and superconducting gap\label{Gap}}
So far, we have discussed the ARPES data and the theoretical consequences in a relatively high-energy region above the order of $J$$\sim$100~meV.  This section is devoted to low energy features of the ARPES spectra including the superconducting gaps and pseudogaps.

\subsection{Experiments\label{GapExp}}
Rapid improvement of energy and momentum resolutions in ARPES made it possible to measure the superconducting-gap anisotropy \cite{Shen1}.  Measurements on Bi2212 below $T_c$ demonstrated an anisotropic gap with the $d_{x^2-y^2}$ symmetry, by plotting the midpoint energy of the leading edge of about 20~meV in energy distribution curves as a function of $0.5|\cos k_x-\cos k_y|$.   In the underdoped samples, the leading edge midpoint shifts were observed not only in the superconducting state but also in the normal state \cite{Loeser1,Ding4}.  As an example, \fref{LeadingEdgeGap} shows the leading edge gaps for two underdoped samples of Dy-doped Bi2212 films \cite{Harris1}.  The midpoint shifts in the normal state almost follow those in the superconducting state.  The deviation from the linearity near the origin in the figure, i.e., near the $d$-wave node position, might be attributed to the `dirty $d$-wave' picture \cite{Harris1}.  The doping dependence of the superconducting and normal state gaps are summarized in \fref{PhaseDiagram} \cite{WhitePJ1}.  In the underdoped region, the values of both gaps are similar to each other with approximately 25~meV, while in the overdoped region the superconducting gap is finite but the normal one is almost zero.  In the phase diagram, the high energy structure at around ($\pi$,0) is also depicted.  This feature corresponds to broad peaks identified in \sref{Insulator} as the high energy pseudogap showing a $d$-wave dispersion [see \fref{CCOCdwave}(a)].  It is interesting that the high energy gap scales with the low energy ones of about 25~meV, i.e., the superconducting gap and low energy pseudogap \cite{Campuzano2}.

The onset temperature of the pseudogap $T^*$ decreases with increasing doping and becomes almost the same as $T_c$ at optimal doping \cite{Ding4}.  In the temperature region between $T_c$ and $T^*$ in the underdoped samples, the Fermi surface shows unusual behavior as schematically illustrated in \fref{FermiArcs} \cite{Norman1}.  Since the pseudogap opens up only at around ($\pi$,0) just below $T^*$, the Fermi surface breaks into disconnected arcs, which gradually shrinks with decreasing temperature from $T^*$.  At $T_c$, a point node expected from $d_{x^2-y^2}$ symmetry emerges.  This clearly demonstrates that the behavior near the ($\pi$,0) point is a clue of the physics of the pseudogap.  In contrast, in the overdoped case the gap opens at the same temperature for all Fermi surface momenta.  The three panels in \fref{FermiArcs} also correspond to the evolution of the Fermi surface from under, optimal to overdoped cases at a fixed temperature \cite{Marshall,Norman1,Norman2}.

In the underdoped case, the ($\pi$,0) point is found to be special when one analyses the temperature and momentum dependence of the gap values.  Norman {\it et al.} \cite{Norman1,Norman2} found that the pseudogap near ($\pi$,0) is independent of temperature in magnitude but is filled in by the spectral weight, while away from ($\pi$,0) the gap closes with the reduction of its magnitude.  \Fref{TempGap} \cite{Norman2} illustrates this behavior.  In the figure, the effect of the Fermi function $f(\omega)$ in \eref{Ikw} is eliminated by plotting the symmetrized intensity $I({\bf k_F},\omega)+I({\bf k_F},-\omega)$.  Analyzing the intensity with the help of phenomenological forms of the self-energy, they obtained the temperature dependence of the gap having the anomalous features mentioned above.  This implies that the temperature dependence of the gap near the node region may be treated by a mean field fashion but not near the ($\pi$,0) region.

\Fref{SuperAkw} shows the ARPES data above and below $T_c$ for a slightly overdoped Bi2212 sample \cite{Norman3}.  Below $T_c$, it shows an interesting behavior near the ($\pi$,0) point:  A sharp peak, for example, at about 40~meV for the curve of M point in \fref{SuperAkw}(b), is followed by a dip at about 80~meV and then by a hump at approximately 120~meV.  From high resolution measurements, the sharp peak is found to have an intrinsic width of approximately 14~meV \cite{Fedorov}.  \Fref{PeakHump} shows a summary of the peak and hump positions deduced from \fref{SuperAkw} \cite{Norman3}.  The sharp peak remains at low energy region independent of momentum, while the hump follows the normal state broad peak position whose momentum dependence is shown in \fref{DopingDispersion}.  Similar features are seen even in the underdoped samples \cite{Campuzano2}.  At present, the peak/dip/hump structure has been observed only for Bi2212 system but not for other systems, Bi2201, YBCO, LSCO and NCCO together with Zn-doped Bi2212 \cite{WhitePJ2}.  This mystery will be resolved in the future by the improvement of energy resolution as well as sample quality.

\subsection{Theories}

Although the $t$-$t'$-$t''$-$J$ model explains the systematic evolution of the line shape in the normal state from undoped to overdoped cases as discussed in \sref{Insulator} and \sref{Doping}, the low energy features related to the superconducting and pseudogaps have not been explained as far as the authors know.  Instead, for the $t$-$J$ model without $t'$ and $t''$, the gauge theories based on the spin-charge separation have been applied and some of the feature are explained.  The energy scale of the dispersion and a gap feature near the ($\pi$,0) point obtained by the U(1) \cite{DaiX} and SU(2) \cite{Wen} theories seems to be consistent with the pseudogap phenomena (see \fref{U(1)+SU(2)}).  Both theories predict small Fermi surface segments at low doping that continuously evolve into the large Fermi surface at high doping, which is consistent with the ARPES data discussed above.  However, the spectral functions in the superconducting state have not been calculated based on the gauge theories.  This remains to be solved in the near future.

Fermi arcs and patches, the latter of which corresponds to a region near ($\pi$,0) where gaps open, were discussed by Furukawa {\it et al.} \cite{Furukawa}.  They used a renormalization group approach for a two-dimensional Fermi liquid near half filling.  Umklapp scattering leads to the instability of the Fermi liquid where a spin and charge gap opens in the patch region.  Engelbrecht {\it et al.} \cite{Engelbrecht} obtained a similar arc-patch behavior, where they consider a two-dimensional model whose ground state is a $d_{x^2-y^2}$ superconductor and show the destruction of the Fermi surface above $T_c$ as a result of pairing correlations.  Both studies have some similarities to a phenomenological arguments proposed by Geshkenbein {\it et al.} \cite{Geshkenbein} where bosonic pairs around the patch region interact with fermions in the arc region.  The breakdown of the Fermi surface satisfying the Luttinger theorem was indicated by the high temperature series expansion for the momentum distribution function of the two-dimensional $t$-$J$ model \cite{Putikka}.

Chubukov and Morr \cite{Chubukov2} studied the peak/dip/hump feature by using a spin-fermion model for cuprates.  Soon after, Abanov and Chubukov \cite{Abanov} associated the peak-dip separation with the resonance peak observed in neutron scattering experiments on Bi2212 \cite{Fong1} as well as YBCO \cite{DaiP} as a result of strong spin-fermion interaction that couples the fermions near the ($\pi$,0) points to the antiferromagnetic interaction with momentum {\bf Q}=($\pi$,$\pi$).  The same scenarios but qualitative arguments that a collective mode peaked at {\bf Q} accounts for the peak/dip/hump structure near ($\pi$,0) have been given by, for example, Shen and Schrieffer \cite{Shen2} and Norman and Ding \cite{Norman4}.  The latter authors also pointed out a fact that the resonance neutron peak position is the same as the peak-dip separation.  This feature was confirmed experimentally by comparing the doping dependence of both quantities \cite{Campuzano2}.

In addition to the models mentioned above, there are many models that propose the mechanism of the pseudogap.  For example, Emery {\it et al.} \cite{Emery1} developed a model based on stripes.  The metallic stripes obtain a spin gap through the pair hopping of electrons between the stripe and adjacent spin domain, resulting in singlet-pair formation in the stripes.  Not only the pair hopping but also spin exchange interaction \cite{Granath} works as the spin gap formation.  Several models for the pseudogap are based on precursor superconductivity as reviewed in \cite{Maly}.  For any proposed models, explicit calculations are required of the spectral functions that satisfactorily explain the ARPES data.

\section{Summary\label{Summary}}
The ARPES spectra and the underlying electronic states of high $T_c$ cuprates have been reviewed from theoretical point of view.  Much emphasis has been put on systematic change of the spectral weight with doping, in particular, near the momentum ($\pi$,0).  We have also shown that various interesting features occur around this momentum.

In the insulator, the dispersion of a doped hole has a close resemblance to a $d$-wave dispersion along the ($\pi$/2,$\pi$/2)-($\pi$,0) direction.  This dispersion is continuously evolved into a normal state dispersion in the doped cases.  At the same time, the ($\pi$,0) spectrum gradually changes from broad structures in the insulating and underdoped cases to a sharp peak structure in the overdoped ones.  These systematics are explained by a $t$-$J$ model with long-range hoppings, $t'$ and $t''$.  Spin-charge separation scenarios also give an alternative explanation for the systematics.  Both approaches, however, seem to share the same underlying physics in a sense that antiferromagnetic interactions play a crucial role.

The dependence of the Fermi surface on temperature and doping clearly demonstrates the anomaly of the normal state in the underdoped system.  This evolution of the Fermi surface to the regions of the Fermi arcs and patches is related to the physics at the ($\pi$,0) point.  The peak/dip/hump feature in the superconducting state is also clearly seen near the ($\pi$,0) point, implying again the anomalous feature of electrons with {\bf k}=($\pi$,0).  Since the hump is related to the normal-state peak, in which strong magnetic interaction in the spin background plays a crucial role, the problem to be clarified may be the interplay of magnetic interaction and superconducting coherence that enhances the peak and dip structures in the superconducting state.  In this context, it is interesting that recent development of the ARPES experiment makes it possible to perform detailed study of the quasiparticle properties along not only the ($\pi$,0) direction but also the ($\pi$,$\pi$) direction \cite{Valla,Kaminski}.  These experiments will provide an opportunity to test the existing theories.

The evidence of the stripes is accumulating, at least, for the underdoped LSCO system.  The broad ARPES spectra along the (0,0)-($\pi$,$\pi$) direction have been shown to be attributable to the presence of vertical stripes (see \sref{LSCO}).  Then, an arising question is whether the stripe picture can be extended to other materials.  There is an argument that flat dispersions along the (0,0)-($\pi$,0) direction and flat segments of the Fermi surface along the direction might be indications of the presence of the stripes \cite{Feng,Balatsky}.  This is based on the fact that Nd-doped LSCO, in which the (static) stripes are known to exist, shows the same flat features \cite{Zhou}.  The relation between the stripes and these flat features remains to be clarified in the near future.

Very recently, it has been pointed out that the spectral weight strongly depends on the incident energy $h\nu$ \cite{Chuang}.  This might suggest that the matrix element given in \eref{SpinonDispersion} depends on energy \cite{Bansil2}.  Further experimental and theoretical investigations of the APRES spectra are necessary.

\ack
We would like to thank Z-X Shen, C Kim, Y Shibata, S Nagai and N Nagaosa for collaboration and discussion on this subject.  We also thank A Ino, K Kobayashi, T Mizokawa, H Eisaki, A Fujimori and S. Uchida for fruitful discussions.  This work was supported by a Grant-in-Aid for Scientific Research from the Ministry of Education, Science, Sports and Culture of Japan, CREST, and NEDO.

\vspace{2cm}
\centerline{Figure Captions}
\vspace{1cm}

\begin{figure}[h]
\caption{Fermi surface determined by ARPES. (a) Bi2212 for heavily underdoped ($T_c$=15~K), slightly underdoped ($T_c$=83~K) and slightly overdoped ($T_c$=87~K) samples.  Solid lines are Fermi surfaces obtained by the tight binding model for different doping.  After Ding {\it et al.} \cite{Ding2}.  (b) La$_{2-x}$Sr$_x$CuO$_4$ for $x$=0.1 (underdoping) and $x$=0.3 (overdoping).  After Ino {\it et al.} \cite{Ino1}.}
\label{FSexp}
\end{figure}

\begin{figure}[h]
\caption{Fermi surface in the tight binding model at half filling (broken line) and 30\% hole doping (solid line) for (a) Bi2212 with $t'$/$t$=$-$0.34 and $t''$/$t$=0.23 and for (b) LSCO with $t'$/$t$=$-$0.12 and $t''$/$t$=0.08.  These parameter values are used in \cite{Kim1} and \cite{Tohyama2}.}
\label{FStheory}
\end{figure}

\begin{figure}[h]
\caption{ARPES data for insulating Sr$_2$CuO$_2$Cl$_2$ at $T$=150~K as a function of kinetic energy of outgoing electrons.  (a) The (0,0)-($\pi$,$\pi$) direction.  (b) The (0,0)-($\pi$,0) direction.  (c) The ($\pi$/2,$\pi$/2)-(2$\pi$/3,0) arc along which the peaks show the highest intensity.  The ticks denote two structures obtained by curve fitting.  After Kim {\it et al.} \cite{Kim1}.}
\label{SCOCspectrum}
\end{figure}

\begin{figure}[h]
\caption{Energy dispersion of quasiparticle for insulating Sr$_2$CuO$_2$Cl$_2$ measured from the top of the band.  Experimental data are taken from \cite{Wells} (open circles), \cite{LaRosa} (open triangles) and \cite{Kim1} (open squares).  Solid circles: the results of the self-consistent Born approximation (SCBA) for the $t$-$t'$-$t''$-$J$ model with $t$=0.35~eV, $t'$=$-$0.12~eV, $t''$=0.08~eV and $J$=0.14~eV.  The solid lines are obtained by fitting the SCBA data to a dispersion relation given by $E_0({\bf k})+E_1({\bf k})$ (see \eref{tJDispersion} and \eref{t't''Dispersion}), being $t'_{eff}$=$-$0.038~eV and $t''_{eff}$=0.022~eV.  The broken line along the ($\pi$,0)-(0,$\pi$) direction represents the spinon dispersion given by \eref{SpinonDispersion} from \cite{Laughlin}.}
\label{SCOCdispersion}
\end{figure}

\newpage
\begin{figure}[ht]
\caption{Doping and carrier dependence of ARPES data of (a) Sr$_2$CuO$_2$Cl$_2$, (b) underdoped and (c) overdoped Bi2212, and (d) Nd$_{1.85}$Ce$_{0.15}$CuO$_4$.  After Kim {\it et al.} \cite{Kim1}.}
\label{ARPESweight}
\end{figure}

\begin{figure}[ht]
\caption{Electron-removal spectral functions at half filling.  (a) $t$-$J$ model and (b) $t$-$t'$-$t''$-$J$ model.  The calculation was performed on a 4$\times$4 cluster, and the $\delta$ functions were convoluted with a Lorentzian broadening of 0.10~eV.  The binding energy is measured from the energy of the first ionization state in the photoemission process.  After Kim {\it et al.} \cite{Kim1}.}
\label{Akw1hole}
\end{figure}

\begin{figure}[ht]
\caption{(a) High energy gap against 0.5$|\cos k_x- \cos k_y|$ plot for Ca$_2$CuO$_2$Cl$_2$ and Bi2212 with various Dy dopings.  (b) Doping dependence of the ($\pi$,0) spectra.  The arrows show the broad peak positions that were assigned as the high energy pseudogap.  After Ronning {\it et al}. \cite{Ronning}.}
\label{CCOCdwave}
\end{figure}

\begin{figure}[ht]
\caption{(a) The dependence of the energy difference $E(\pi,0)-E(\pi/2,\pi/2)$ on the values of $t'$ and $t''$.  The variable $\alpha$ represents a scaling
factor of $t'$ and $t''$ as $t'(t'')=\alpha t'_0(t''_0)$, being $t'_0$=$-$0.34$t$ and $t''$=0.23$t$.  Solid line: spin-density wave mean-filed result of the $t$-$t'$-$t''$ Hubbard model.   Solid circles: self-consistent Born approximation of the $t$-$t'$-$t''$-$J$ model on a 16$\times$16 lattice with $J$/$t$=0.4.  (b) $\left(E({\bf k})-E(\pi/2,\pi/2)\right)$ versus $J$/$t$ in the case of $\alpha$=1.  The solid line represents 2.2$J$/$t$ known as the band width from ($\pi$/2,$\pi$/2) to (0,0).  After Tohyama {\it et al.} \cite{Tohyama1}.}
\label{BandWidth}
\end{figure}

\begin{figure}[ht]
\caption{Electron-removal spectral function $A({\bf k},\omega)$ at half filling, and spin correlation $S({\bf q},\omega)$ and charge correlation $N({\bf q},\omega)$ with {\bf q}={\bf k}-{\bf k}$_0$ at one-hole doping for a $\sqrt{20}$$\times$$\sqrt{20}$ $t$-$t'$-$t''$-$J$ model.  $J$/$t$=0.4.  {\bf k}=($\pi$,0) and {\bf k}$_0$=(3$\pi$/5,$-\pi$/5).  (a) $t$-$J$ and (b) $t$-$t'$-$t''$-$J$ with realistic values of $t'$ and $t''$.  The quasiparticle peak in the $t$-$J$ model carries both spin and charge degrees of freedom, but not for the $t$-$t'$-$t''$-$J$ model.  The insets in (a) and (b) show $A({\bf k},\omega)$ with full energy scale.  After Tohyama {\it et al.} \cite{Tohyama1}.}
\label{AkwSqwNqw}
\end{figure}

\begin{figure}[ht]
\caption{Doping dependence of the dispersion of peak centroids for Bi2212 with various doping levels.  Optimal and overdoped samples show a cut at the Fermi level along the ($\pi$,0)-($\pi$,$\pi$) line.  An insulating sample does not show a cut along the (0,0)-($\pi$,$\pi$) line.  The unit of energy is eV.  After Marshall {\it et al}. \cite{Marshall}.}
\label{DopingDispersion}
\end{figure}

\newpage
\begin{figure}[ht]
\caption{Electron-removal spectral functions for various dopings in the $t$- $t'$- $t''$-$J$ model. (a) Hole-doping case with $\delta$=0.1, (b) hole-doping case with $\delta$=0.3, and (c) electron doping with $\delta$=0.2.  The calculation was performed on a $\sqrt{20}$$\times$$\sqrt{20}$ cluster, and a Lorentzian broadening of 0.10~eV was used.  The binding energy is measured from the energy of the first ionization state in the photoemission process.  The quasiparticle peak at ($\pi$/5,3$\pi$/5) for of $\delta$=0.3 in (b) appears above the Fermi level.  After Kim {\it et al.} \cite{Kim1}}
\label{Akwdoping}
\end{figure}

\begin{figure}[ht]
\caption{(a) Quantum Monte Carlo study of spectral functions of a Hubbard model with $U$/$t$=10 and $t'$/$t$=$-$0.35 at doping $\delta$=0.06 and temperature $T$=$t$/3 on a 6$\times$6 cluster.  The ($\pi$,0) spectrum is very broad.   (b) The energies of the dominant peaks in the spectral functions.   The full circles form the quasiparticle band, and their diameter is proportional to the intensity.  The squares correspond to the incoherent part of the spectrum.  After Duffy {\it et al.} \cite{Duffy}.}
\label{QMC094}
\end{figure}

\begin{figure}[ht]
\caption{(a) Spectral function of the $t$-$J$ model obtained by using the U(1) gauge theory with staggered gauge field fluctuation.  Left panel: underdoping with $\delta$=0.09. {\bf k}=($\pi$/2,$\pi$/2) for solid line and ($\pi$,0) for dashed line. Right panel: overdoping with $\delta$=0.20.  {\bf k}=(0.45$\pi$,0.45$\pi$) for solid line and ($\pi$,0) for dashed line.  After Dai and Su \cite{DaiX}.  (b) Quasiparticle dispersions of the $t$-$J$ model in the staggered flux phase of the SU(2) theory.  The vertical bars are proportional to the peak intensity.  After Lee {\it et al.} \cite{Wen}.}
\label{U(1)+SU(2)}
\end{figure}

\begin{figure}[ht]
\caption{ARPES spectra of underdoped ($x$=0.1) and optimally doped ($x$=0.15) La$_{2-x}$Sr$_x$CuO$_4$.  Insets show measured points and the polarization of incident photons (arrows).  After Ino {\it et al.} \cite{Ino1}.}
\label{LSCOspectrum}
\end{figure}

\begin{figure}[ht]
\caption{Spectral function $A({\bf k},\omega)$ of the $t$-$t'$-$t''$-$J$ model with the stripe potential on a $\sqrt{18}$$\times$$\sqrt{18}$ two-hole cluster.  Parameter values are those for LSCO.  (a), (b), and (c) are results for the potential of $V$/$t$=0, 1, and 2, respectively.   The solid and dashed curves for the electron removal and addition spectra, respectively, are obtained by a Lorentzian broadening.  The momentum is measured in units of $\pi$.  (d) The density of states obtained by $\sum_{\bf k} A({\bf k},\omega)$.  After Tohyama {\it et al.}  \cite{Tohyama2}.}
\label{ARPESstripe}
\end{figure}

\begin{figure}[ht]
\caption{Spectral function of a stripe model containing disordered charge stripes and antiphase spin domains.  (a) The spectral density in the vicinity of the Fermi level.  (b) The dispersion relation, and (c) the corresponding spectral density of the dispersion.  After Salkola {\it et al.} \cite{Salkola}.}
\label{StripeBand}
\end{figure}

\begin{figure}[ht]
\caption{Leading edge midpoint shifts from the Fermi level in the superconducting and normal states of Dy-doped Bi2212 samples.  A $d_{x^2-y^2}$ gap would be a straight line on this plot.  After Harris {\it et al.} \cite{Harris1}.}
\label{LeadingEdgeGap}
\end{figure}

\begin{figure}[ht]
\caption{Phase diagram for the doping dependence of the energy gaps at ($\pi$,0) in Bi2212.  The gap values of the normal state (NG) and of the superconducting state (SG) are almost the same in the underdoped region.  The shaded area at higher energy corresponds to the broad peak structure shown in \fref{CCOCdwave}.  After White {\it et al.} \cite{WhitePJ1}.}
\label{PhaseDiagram}
\end{figure}

\begin{figure}[ht]
\caption{The dependence of the Fermi surface on temperature in the underdoped samples.  The $d_{x^2-y^2}$-type node below $T_c$ (left panel) becomes a gapless arc above $T_c$ (middle panel) and then becomes the full Fermi surface at $T^*$ (right panel).  The evolution also corresponds to the doping dependence of the Fermi surface from underdoped to overdoped samples at a fixed temperature.  After Norman {\it et al.} \cite{Norman1}.}
\label{FermiArcs}
\end{figure}

\begin{figure}[ht]
\caption{The temperature dependence of energy gaps at different momenta in the underdoped Bi2212.  (a) Symmetrized ARPES spectra at (open circles) {\bf k}$_F$ point 1 in the zone, and at (open triangles) {\bf k}$_F$ point 2.  A phenomenological model is used to fit the data (solid lines).  (b) The gap function $\Delta(T)$ obtained by the fitting for the two {\bf k$_F$} points.  After Norman {\it et al.} \cite{Norman2}.}
\label{TempGap}
\end{figure}

\newpage
\begin{figure}[ht]
\caption{ARPES data for slightly overdoped Bi2212.  (a) The normal state, and (b) and (c) the superconducting state.  In the superconducting state, the peak/dip/hump feature is seen at around the ($\pi$,0) point.  After Norman {\it et al.} \cite{Norman3}.}
\label{SuperAkw}
\end{figure}

\begin{figure}[ht]
\caption{Positions of the sharp peak and the broad hump in the superconducting state of Bi2212, which are obtained from \fref{SuperAkw}.  The hump position is proportional to that of the normal state peak, while the peak position is almost constant.  After Norman {\it et al.} \cite{Norman3}.}
\label{PeakHump}
\end{figure}

\end{document}